\documentclass[twocolumn,showpacs,preprintnumbers,amsmath,amssymb]{revtex4}

\usepackage{graphicx}
\usepackage{dcolumn}
\usepackage{bm}

\begin{document}


\title{Magnetically-induced ferroelectricity in orthorhombic manganites: \\ microscopic origin and chemical trends}

\author{Kunihiko Yamauchi$^1$}
\author{Frank Freimuth$^2$}
 \author{Stefan Bl\"{u}gel$^2$}
\author{Silvia Picozzi$^1$}%
 \email{silvia.picozzi@aquila.infn.it}
\affiliation{%
1. Consiglio Nazionale delle Ricerche - Instituto Nazionale di Fisica della Materia (CNR-INFM), CASTI Regional Lab., 67100 L'Aquila, Italy\\
2. Institut f\"ur Festk\"orperforschung, Forschungszentrum J\"ulich, 52425 J\"ulich, Germany
}%

\date{\today}

\begin{abstract}

The microscopic origin of the magnetically-driven ferroelectricity in collinear AFM-E orthorhombic manganites is explained by means of  first-principles Wannier functions. We show that
the polarization is mainly determined by the {\em asymmetric} electron hopping of orbitally-polarized $e_g$ states, implicit in the peculiar 
in-plane zig-zag spin arrangement in the AFM-E configuration. In ortho-HoMnO$_3$, Wannier function centers are largely displaced with respect to corresponding ionic positions, implying that the final polarization is strongly affected by a purely electronic contribution, at variance with standard ferroelectrics where the 
ionic-displacement is dominant. However,  the final value of the polarization is the result  of competing effects, as shown by the opposite signs of  the 
contributions to the polarization coming from  the Mn
$e_g$ and $t_{2g}$ states. Furthermore, a systematic analysis of the link between ferroelectricity and the spin, orbital and lattice degrees of freedom in the manganite series has been carried out, in the aim of ascertaining chemical trends as a function of the rare-earth ion. Our results show that the Mn-O-Mn angle is the key quantity in determining the exchange coupling: upon decreasing the Mn-O-Mn angle,
 the first- (second-) nearest neighbor ferromagnetic (antiferromagnetic) interaction decreases (remains constant), in turn stabilizing either the AFM-A or the AFM-E spin configuration for weakly or strongly distorted manganites, respectively. The Mn $e_g$ contribution to the polarization dramatically increases with the Mn-O-Mn angle
 and decreases with the ``long" Mn-O bond length, whereas the Mn $t_{2g}$ contribution decreases with the ``short" Mn-O bond-length, partially cancelling the former term. 

\end{abstract}

\pacs{75.47.Lx,75.80.+q, 75.50.Ee,  77.80.-e }

\maketitle

\section{\label{sec:intro}Introduction\protect\\}

Multiferroics are attractive multifunctional materials where magnetism and ferroelectricity coexist and are generally coupled.
In particular, orthorhombic rare-earth manganites $R$MnO$_{3}$ represent an important class of  ``improper ferroelectrics'' \cite{mostche} where electric dipoles are induced by a frustrated magnetic ordering.
Within this family of compounds, TbMnO$_{3}$ and DyMnO$_{3}$, in their noncollinear magnetic phases,
have been experimentally shown to behave as multiferroics showing a weak polarization
($P<0.1\mu C/cm^{2}$)
and a spin-flop ferroelectric transition.\cite{kimuranat} 
Moreover, it has been recently predicted that
relatively strong ferroelectricity occurs in the E-type antiferromagnetic (AFM) phase of $R$MnO$_{3}$ through a model study 
where the double-exchange interaction between Mn-$d$ orbitals is proposed as a driving force for polar atomic displacements.\cite{ivan} In that case, the electric polarization $P$ is not related to Dzyaloshinskii-Moriya interaction, which is expected to generate much lower $P$.\cite{mostovoy,ivansoc,nagaosa}
Following the model study, our previously reported ab-initio calculations \cite{prlslv} have confirmed that 
AFM-E HoMnO$_{3}$ indeed shows a high ferroelectric polarization ($P\sim 6\mu C/cm^{2}$); it was shown there that the AFM-E spin ordering, which breaks the space-inversion symmetry, is such that ferroelectric dipoles arise even without atomic displacements.\cite{prlslv} From the experimental point of view, ferroelectricity was detected in policrystalline ortho-HoMnO$_3$,\cite{lorenz} the magnitude of polarization being however much smaller than what is theoretically predicted and showing a strong dependence on the magnetic field below the ordering temperature of the Ho, suggesting their involvement in the development of $P$. The reason behind the disagreement between theory \cite{ivan,prlslv} and experiments \cite{lorenz} is still under debate.

The microscopic origin of the multiferroism in ortho-$R$MnO$_{3}$ is tightly linked to the lattice degree of freedom: a small ionic radius of the $R$ atom holds a key role in affecting {\em i})
 structural properties, such as both the JT distortion 
in the MnO$_{2}$ plane and the GdFeO$_{3}$-like tilting of MnO$_{6}$ octahedron,
{\em ii}) spin configuration determined by Mn-$d$ super-exchange interaction $J_{ij}$, 
{\em iii}) orbital ordering stabilized by Jahn-Teller (JT) distortion, %
and {\em iv}) hopping integral in terms of double exchange interaction.
Since these aspects are all mutually combined, 
first we will discuss the case of AFM-E HoMnO$_3$, and we will explain, from a microscopic quantum-mechanical point of view, the origin of the magnetically-induced ferroelectric polarization in terms of the spin, orbital and lattice degrees of freedom. 
The analysis based on Wannier functions represents a novel development with respect to our previous study on HoMnO$_{3}$.\cite{prlslv}
Moreover, in the second part of the paper, we will discuss about structure, magnetism, orbital ordering, hopping integrals and ferroelectricity - as well as the links between them - along the manganites series, with the main aim of identifying chemical trends as a function of the $R$ ion.

\section{\label{sec:comp}Structural and Computational Details\protect\\}

The unit cell in orthorhombic $R$MnO$_{3}$ 
shows strong distortions with respect to the ideal cubic perovskite. Whereas the $Pbnm$ setting is used in some other references, the standard $Pnma$ orientation is adopted in this paper, {\em i.e.}
 we choose $b$ as the longest axis. 
Although the hexagonal non-perovskite phase is more stable than the orhorhombic phase
in $R$MnO$_{3}$ for $R$ smaller than Tb,
the transition to the metastable orthorhombic phase can be obtained by high-pressure synthesis.\cite{struc_zhou,struc_tachibana,struc_mori,struc_alonso}
We have considered five kinds of AFM configurations:
AFM-A, C, G in 20 atoms/unit cell and AFM-E and AFM-E* in 40 atoms/unit cell. 
We recall that, according to the standard Wollan-Koehler notation\cite{wollan}, the AFM-A shows FM (AFM) intraplanar (interplanar) coupling; the AFM-C shows FM (AFM) interplanar (intraplanar) coupling; the AFM-G shows AFM in both intra- and interplanar coupling. The AFM-E  shows in-plane FM zig-zag chains antiferromagnetically coupled to the neighboring chains; the interplanar coupling is also AFM
(for the AFM-E spin arrangements in the MnO$_2$ plane, we refer to the following Fig. \ref{fig:afm}).  We denote by AFM-E* the spin-configuration showing the same in-plane spin arrangement  as AFM-E,  but  with an interplanar FM coupling. 
Note that the AFM-A spin arrangement shows the space-inversion as symmetry operation, 
at variance with the AFM-E spin-configuration which shows non centrosymmetric $Pmn2_{1}$ symmetry. 
 
 As far as the electronic structure is concerned (especially for the density of states projected on the Mn atom as well as orbital-ordering, see below), it is useful to define  - in addition to the ``global" $X, Y, Z$ orthorhombic frame - a ``local" frame, specific to each Jahn-Teller-like distorted MnO$_6$ octahedron, obtained by choosing $x, y, z$ along the middle, short and long Mn-O axis, respectively.
In this local frame, the orbital ordered Mn-$e_{g}^{1}$ state, which is often expressed as $(3x^{2}-r^{2})/(3y^{2}-r^{2})$, is described as $3z^{2}-r^{2}$ (denoted as $z^{2}$ hereafter for simplicity).
The simulations were performed by using
density-functional theory (DFT) and the Perdew-Becke-Erzenhof (PBE) version of the generalized gradient approximation (GGA) to the exchange-correlation potential.\cite{pbe}
The calculations were done with two program codes according to different purposes.

The expensive calculations for the structural optimization of the atomic structure and Berry phase for the AFM-A and AFM-E phases
were done with ``Vienna $Ab$ $initio$ Simulation Package (VASP) program code'',\cite{vasp} 
where projector-augmented-wave (PAW) pseudo-potential is used.
The plane wave cutoff is set to 500 eV in energy. We used
8 special k-points (divided as $4\times3\times4$) in 1/8 irreducible Brillouin zone (IBZ) for the A-AFM phase
and 4 special k-points (divided as $2\times3\times4)$ in 1/4 IBZ for the E-AFM phase 
 according to the Monkhorst-Pack scheme.
The Berry phase was calculated by integrating over six k-point strings
 parallel to the c axis, each string containing 6 k-points.

Calculations requiring a higher precision ({\em i.e.} as for 
the total energy differences for the stability of different spin-configuration) and the construction of Maximally-localized Wannier functions (WFs) \cite{marzari}
were done with the FLEUR code\cite{fleur} which is based on the 
full-potential linearized augmented planewave (FLAPW) formalism.\cite{flapw}
Muffin-tin radii were set to 
2.5, 2.0 and 1.5 a.u. for $R$, Mn and O atoms, respectively,
where the wavefunction cutoff was chosen as 3.8 a.u.$^{-1}$.
The potential was converged with 24 special \textbf{k}-points
and density of states was calculated with 192 \textbf{k}-points within the  tetrahedron method.
For AFM-E and AFM-E$^{*}$ phase, 12  \textbf{k}-points were used according to the doubled unit cell. The
Wannier function calculation, whose procedure was recently implemented in the FLEUR code \cite{freimuth}, was done with 512 \textbf{k}-points (divided as $8\times8\times8$).
$R$-$5s$ and $5p$ states were treated as local orbitals.

The localized $R$-$4f$ electrons were assumed as core electrons:
``frozen core'' within the VASP code and ``open core'' within the FLEUR code, \cite{4f}
where the spin moment is maximized due to Hund's rule.
Irrespective of these approaches, the $4f$ states lie deep in energy 
(a few eV below the Fermi energy) and they are almost completely undispersed, so that they do not affect other valence states. 

\section{Microscopic origin of the ferroelectric polarization in H\lowercase{o}M\lowercase{n}O$_3$}

\begin{figure}
\resizebox{68mm}{!}
{
\includegraphics{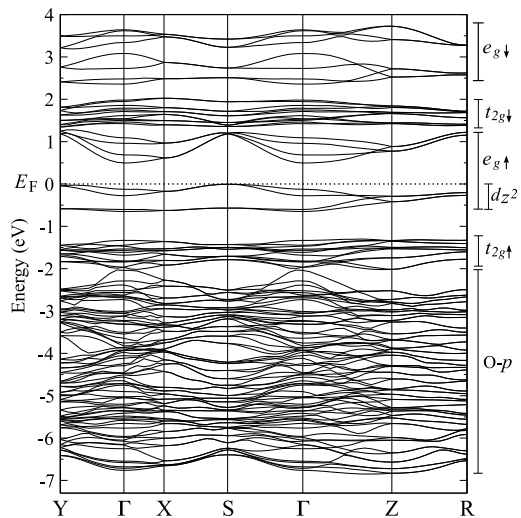}
}
\caption{\label{fig:band} Energy bandstructure of optimized AFM-E HoMnO$_{3}$. Due to the AFM-E state, only the up-spin channel  is plotted. $E_\textrm{F}$ denotes the Fermi energy and it is set as zero of the energy scale.
}
\end{figure}
Let us start our discussion by focusing on the case of AFM-E HoMnO$_3$, which was previously found by  some of us to show a large polarization along the in-plane $c$-axis due to non-centrosymmetric collinear spin-arrangement.\cite{prlslv} In this work, we perform a careful analysis in terms of Wannier functions which explains the microscopic mechanism at the basis of the final ferroelectricity. 
The structural parameters used in our calculations are summarized in a later section, together with the parameters in other $R$MnO$_{3}$ compounds (cfr Table \ref{struc.a} below). 
The magnetically-induced spontaneous polarization was calculated by using {\em i}) the point charge model (PCM) where each ion has been given its nominal charge (Ho:+3, Mn:3+, O:2-); {\em ii}) 
 the Berry phase (BP) method implemented in the VASP code and {\em iii}) the Wannier function (WF) method implemented in the FLEUR code.\cite{notaberry} 
We recall that, in the first approach
only the positions of the anions and cations are considered, whereas in the two latter\cite{berry1,berry2} 
quantum-mechanical treatments, the self-consistent electronic structure is fully taken into account.

\begin{figure}
\resizebox{78mm}{!}
{
\includegraphics[angle=0]{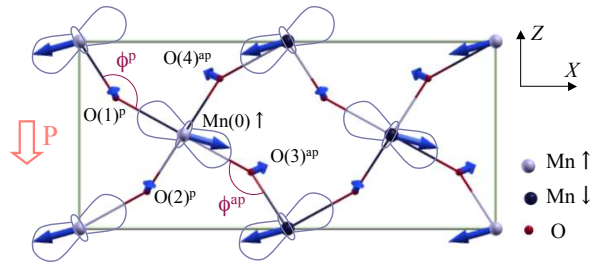}
}
\caption{\label{fig:afm} Atomic displacements in FE HoMnO$_3$, as obtained by the difference of atomic coordinates in optimized AFM-E and optimized AFM-A spin configuration (length of arrows in arbitrary units). A schematic representation of the $e_g$ orbital-ordering and the direction of the ferroelectric polarization are also shown.}
\end{figure}

To calculate the polarization within BP, an adiabatic path from AFM-A to AFM-E phase is assumed, 
in such a way that the direction of Mn spins are progressively rotated from an in-plane FM to a zig-zag-like arrangement.\cite{notaperp} 
The ionic contribution from core electrons and protons are added to BP calculated for the fully occupied valence states.
In Fig.\ref{fig:afm} and Table \ref{tab:disp_atom} we show the displacements of the atoms in the non-centrosymmetric AFM-E spin-configuration with respect to the centrosymmetric AFM-A structure. These quantities will be needed in the discussion reported below.

By using WFs, one can decompose the total polarization into contributions coming from each set of orbitals.
For clarity, we show in Fig.\ref{fig:band} the AFM-E HoMnO$_3$ insulating bandstructure where the relevant states are highlighted: Mn-$e_{g}$, Mn-$t_{2g}$ and O-$p$ orbitals.
We have projected these three groups of occupied eigenstates  into real-space basis separately and ``maximally-localized" them to obtain the corresponding WFs.
The contribution from deeper occupied valence states (such as O-$2s$ and $R$-$5s$, -$5p$ states) is neglected in the WF approach. 
Here, the total polarization  is the sum of the displacements of the centers of each WF 
from the position of the corresponding ion 
plus PCM contribution. 
For the details of the construction and interpretation of WFs, see Ref.\onlinecite{freimuth}.

According to the different approaches, our estimated values for the polarization in AFM-E HoMnO$_3$ are $P_\textrm{BP}$ = -6.14 $\mu$C/cm$^{2}$, $P_\textrm{WF}$ = -5.61 $\mu$C/cm$^{2}$ and $P_\textrm{PCM}$  = -2.09 $\mu$C/cm$^{2}$
 along $Z$ axis.
The values of $P$ along $X$ and $Y$ axis are both negligible in each approach.
The large difference of the values of $P_\textrm{WF}$ ($P_\textrm{BP}$) from $P_{\textrm{PCM}}$ is remarkable.

\subsection{Wannier Representation of Anomalous Contributions} 
The difference between polarization 
$P_\textrm{WF}$ ($P_\textrm{BP}$)
and $P_\textrm{PCM}$
 is referred to as the \textit{anomalous contribution}\cite{gonze} 
 and it is analyzed by means of Wannier functions.
Table \ref{tab:disp_wf} shows the orbital-decomposed polarization obtained by the displacement of the center of WF 
(WFC) with respect to ionic positions for each set of bands.
By definition, the sum of the WFC displacements gives the difference between $P_{\textrm{WF}}$ and $P_{\textrm{PCM}}$, which can be regarded as the difference between the dynamical charge $Z^{*}$ and static (nominal) charge $Z$. In other words, it quantifies the effects of going from a more ionic charge distribution to more covalent bonding state. By means of the WF analysis, not only one can obtain relevant information on which orbital causes the anomalous contribution,  but also a real-space picture of the polar orbital states. 

\begin{table}
\caption{\label{tab:disp_atom}
Atomic displacements from centrosymmetric AFM-A phase in ferroelectric AFM-E HoMnO$_{3}$ (\AA).
O(1)$^\textrm{p}$ and O(2)$^\textrm{p}$ (O(3)$^\textrm{ap}$ and O(4)$^\textrm{ap}$) are connected by $C_{2Z}$ symmetry operation. 
The used notation of atoms is shown in Fig.2.
}
\begin{ruledtabular}
\begin{tabular}{lccccc}
         &$\Delta X$&$\Delta Y$&$\Delta Z$&$|dr|$\\
         \hline
Mn(0)&0.036  &0.003 &-0.010 &0.037\\
O(1)$^\textrm{p}$&-0.006   & 0.009&  0.016&0.020\\
O(3)$^\textrm{ap}$& 0.020   &0.010  &0.009 & 0.024\\
\end{tabular}
\end{ruledtabular}
\end{table}
\begin{table}
\caption{\label{tab:disp_wf}
WFC displacements from atomic coordinates in AFM-E HoMnO$_{3}$ (\AA).
The orbitals are denoted in a local frame ($x$:middle, $y$:short, $z$:long axis for Mn,
$x$:antibonding inter-plane, $y$:antibonding in-plane, $z$:bonding for O).
Only up spin contribution is shown.
}
\begin{ruledtabular}
\begin{tabular}{llrrrrrrrrrrr}
         &&$\Delta X$&$\Delta Y$&$\Delta Z$&$|dr|$\\
         \hline
Mn(0)&$e_{g}:z^{2}$     &-0.169  & 0.044 &0.188 & 0.257\\
         &$t_{2g}:xy$&                     0.021 & -0.008 & -0.017 &   0.028\\
         &$t_{2g}:yz$&                 -0.049 &  0.018 & -0.146 &   0.155\\
         &$t_{2g}:zx$&                     0.013 &  0.007 & -0.008 &   0.016  \\
         \hline
O(1)$^\textrm{p}$&$p_{x}$&-0.158 &  0.071 &  0.200 &   0.265\\
      &$p_{y}$&-0.110 &  0.048 &  0.209 &   0.241\\
      &$p_{z}$& 0.004 &  0.037 &  0.108 &   0.115\\
      \hline
O(3)$^\textrm{ap}$&$p_{x}$& 0.020 & -0.001 & -0.026 &   0.033\\
      &$p_{y}$&-0.025 & -0.013 & -0.038 &   0.047\\
      &$p_{z}$&-0.171 & -0.016 &  0.048 &   0.178\\ 
\end{tabular}
\end{ruledtabular}
\end{table}
\begin{figure}
\resizebox{78mm}{!}
{
\includegraphics[angle=0]{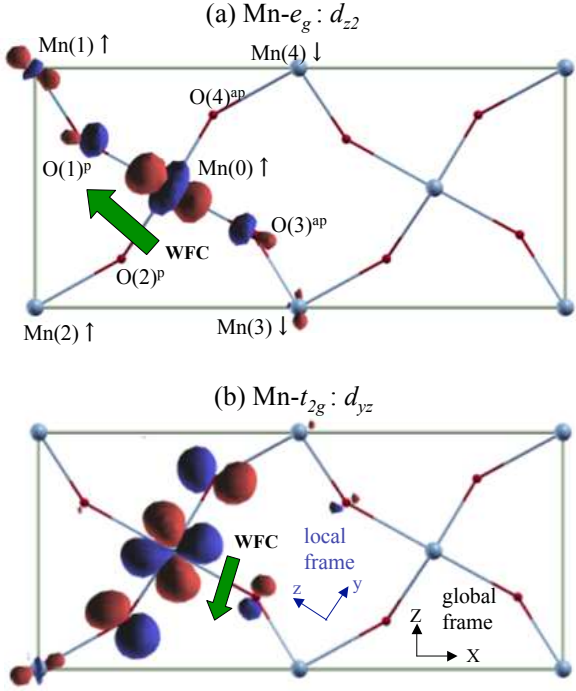}
}
\caption{\label{fig:wan} Isosurface of WFs for Mn-d states centered at Mn(0) site in AFM-E HoMnO$_{3}$: a) $e_g$ states and b) $t_{2g}$ states. Green arrows indicate the displacement of the centers of WFs from the Mn atomic position. Here, the superscript $p$ ($ap$) of O denotes oxygen ion between parallel (antiparallel) spin of Mn ions.}
\end{figure}
\begin{figure}
\resizebox{78mm}{!}
{
\includegraphics[angle=0]{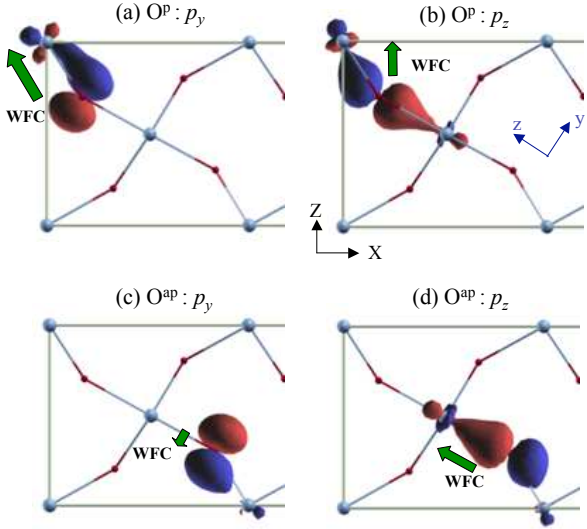}
}
\caption{\label{fig:wan_op} Isosurface of WFs of O-$2p$ state. Only four typical WFs in the $ac$ plane are shown. a) $O^\textrm{p}$ -$p_y$, b) $O^\textrm{p}$-$p_z$, c)$O^\textrm{ap}$-$p_y$, d)$O^\textrm{ap}$-$p_z$ (up spin state). Arrows indicate the displacement of the centers of WFs from the O atomic position. }
\end{figure}
\begin{figure}
\resizebox{78mm}{!}
{
\includegraphics[angle=0]{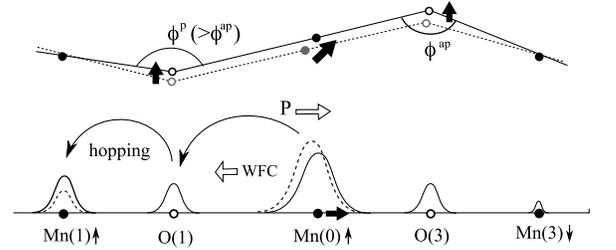}
}
\caption{\label{fig:wan_punch} Schematic representation of the mechanism which causes the microscopic polarization of Mn-$e_{g}$ orbital due to the asymmetric electronic hopping. Bottom sketch: Hopping is allowed only from Mn(0) to Mn(1) and not to Mn(3): this determines the direction of WFC (empty arrow). To increase hopping, Mn(0) moves ``right" (see filled arrow), giving a lower weight of WF  on Mn(0) but larger on Mn(1) [represented as dashed vs solid (before vs after displacement) Gaussians on the atoms]. Also shown is the resulting P (see arrow). Top sketch: Resulting atomic displacement (shown as filled arrows) aimed at increasing $\phi^p$.}
\end{figure}
Let us first focus on the $e_g$-like $d_{z^{2}}$ and $d_{yz}$ orbitals which, together with O-$p$ orbitals, mostly contribute to polarization (see main displacements along the polar $c$ axis inTable. \ref{tab:disp_wf}).

First, consider the  $e_g$-orbital at Mn(0)-up site surrounded by four O ions
and  four second-neighbor Mn ions in the $ac$ plane (Fig.\ref{fig:wan} a).
Because of the JT-derived orbital ordering,
the $e_g$:$d_{z^{2}}$ orbital of Mn(0) ion is pointing towards the two O ions along long axis ($z$ in the ``local" frame).
Therefore, it makes a  $\sigma$ bonding with O(1)$^\textrm{p}$- and O(3)$^\textrm{ap}$-$p_{z}$ orbitals.
In addition, when considering the AFM-E spin configuration,
the up-spin electron of the $e_{g}$-orbital of the Mn(0) ion can hop only onto Mn(1) (up spin) site via the O(1)$^\textrm{p}$ atom but not on the Mn(3) (down spin) site.
This {\em asymmetric} hopping gives the considerable deviation of the WFC from the atom position 
(0.26 \AA) whose direction is shown as a green arrow in Fig.\ref{fig:wan}a. 
Recall that the hopping integral between Mn-$e_{g}$ orbitals 
depends on both the Mn-O bond length $d$ and the Mn-O-Mn angle $\phi$,
 the hopping energy being therefore optimal 
for  larger $\phi$.
As a consequence, in order to increase the hopping, the Mn(0) ion is expected to move toward O(3)$^\textrm{ap}$ to increase $\phi^{p}$ between parallel Mn spins, {\em i.e.} along a direction which is opposite to the WFC displacement.
In the same aim of increasing hopping, O(1)$^\textrm{p}$ moves perpendicularly to the Mn$^{\uparrow}$-Mn$^{\uparrow}$ bonding. 
Similarly, the O(3)$^\textrm{ap}$  atom is displaced perpendicularly to the Mn$^{\uparrow}$-Mn$^{\downarrow}$ bonding (see rough schematization of atomic displacements shown in Fig.\ref{fig:wan_punch}).
 This is consistent with what is shown  in Fig.\ref{fig:afm} and Table \ref{tab:disp_atom}. 
{\em The ions therefore move to enhance the polarization induced by the asymmetric hopping of $e_{g}$-orbital electrons.}
Our proposed mechanism is quantitatively confirmed by the values of the $e_g$ hopping integrals
from the WF centered at Mn(0) site to the one at Mn(1) site (taken from off-diagonal matrix element of the Hamiltonian, see Ref. \cite{freimuth}): 
  in the optimized AFM-E phase, the hopping integral $t=117$ meV is enhanced with respect to $t=100$ meV in optimized AFM-A phase (despite the increase in the long Mn-O bond-length after ionic relaxation which would counteract the increase in hopping given by the increased Mn-O-Mn angle).
 The hopping in the optimized AFM-E phase is also increased with respect to $t$ = 108 meV in  optimized AFM-A atomic coordinates but with the AFM-E spin configuration. 
  In parallel, the band energy ({\em i.e.} on-site energy, taken from diagonal Hamiltonian elements) related to $e_{g}$ states
  reduces along the AFM-A$\rightarrow$AFM-E adiabatic path with the ionic displacements, being minimal in the optimized AFM-E phase ({\em i.e.} 8.98 eV vs 9.03 eV). 
 Therefore, the increase in $t$ and decrease in energy in the AFM-E spin-configuration is mainly determined by the Mn-O-Mn angle dependence and to a less extent by the Mn-O bond-length.
   
We further remark that the movement of the Mn ion (cfr Fig. \ref{fig:wan_punch}) causes a current whose direction coincides with the current by the electron hopping and, therefore, it reinforces the net electronic polarization.
It should be noted that the atomic displacement (0.04 \AA) is very small 
compared to the deviation of the Wannier center (0.26 \AA).
We recall in fact that the atomic displacement is just a secondary effect which occurs in order to enhance the
asymmetric hopping integrals. This magnetically-induced mechanism is therefore different from  the conventional interpretation of polarization  in standard ferroelectrics\cite{gonze}
where the atomic displacement dominates the effect. 

Next, consider the $t_{2g}$:$d_{yz}$ orbital which makes a $\pi$-like bonding with surrounding oxygen $p$ orbitals.
Because the $d_{yz}$ orbital has isotropic symmetry in the plane,
the hybridization with the $p$ orbital depends only on the bond-distance.
In Fig.\ref{fig:wan}b, strong hybridization of $d_{yz}$ orbital with O(2)$^\textrm{p}$-$p_{z}$ and O(4)$^\textrm{ap}$-$p_{z}$ 
is shown. 
Similar to the $e_{g}$ orbital case, the electron hops only into Mn(2)-up site 
so that the WFC is displaced in such a way (see green arrow in Fig.\ref{fig:wan}b).
Moreover, the atomic displacement induced by the $e_g$ orbital  - explained above -
causes  a shorter bond length between Mn(0) and O(3)$^\textrm{ap}$ ions (see weight of Mn-$t_{2g}$ WF on O(3)$^\textrm{ap}$); as a consequence,
the increased hybridization slightly changes the direction of the WFC displacement with respect to the Mn(0)-Mn(2) direction.
Therefore, in HoMnO$_{3}$, 
the anomalous contributions from Mn-$e_{g}$ and $t_{2g}$ orbitals almost cancel each other along the polar $c$ direction, whereas
the O-$p$ contribution survives. 
In order to complete our analysis, we therefore show in Fig. \ref{fig:wan_op} the $O_p$ WF. What happens is qualitatively explained as follows: {\em i})
the  WFC of O$^\textrm{p}$:$p_{y}$ state (forming a $\pi$-bonding with the same-spin Mn-$t_{2g}$ state) is pulled by the short-distant Mn(1) ion (cfr Fig. \ref{fig:wan_op} a);
 {\em ii}) the WFC of O$^\textrm{ap}$:$p_{z}$ which makes a $\sigma$-bonding with the Mn(0)-$e_{g}$ state is pulled by the Mn(0) ion in the same way (cfr Fig. \ref{fig:wan_op} d);
{\em iii}) the WFC of O$^\textrm{p}$:$p_{z}$ which forms a $\sigma$-bonding with both Mn(1)- and Mn(0)-$e_{g}$ states is displaced towards the Mn(1) ion but also moves to increase the Mn-O-Mn angle, with a resulting displacement roughly parallel to the $c$ axis (cfr Fig. \ref{fig:wan_op} b);
{\em iv})  the WFC of O$^\textrm{ap}$:$p_{y}$ which does not form strong bonding with short-distant Mn ions shows a small displacement (cfr Fig. \ref{fig:wan_op} c).

\section{Chemical trends in \textit{R}M\lowercase{n}O$_{3}$}

From the experimental point of view, the magnetic trend of \textit{R}MnO$_{3}$  was found to be strongly affected by the Mn-O-Mn bond angle $\phi$ in the MnO$_{2}$ plane.
From \textit{R}=La to Gd, where $\phi$=155-146$^\circ$, the ground state is A-type AFM; upon decreasing of $\phi$, 
there is an intermediate  ``lock-in incommensulate-AF state'' which couples to a ferroelectric polarization at $R$=Tb and Dy, followed by a transition to the E-type AFM  observed from \textit{R}=Ho to Lu, where $\phi$=144-140$^\circ$.\cite{kimuraprb}

When $\phi$ is close enough to 180$^\circ$, the in-plane FM coupling in AFM-A phase
can be explained on the basis of Kanamori-Goodenough rules \cite{kanamorigoodenough}   in the framework of Mn-O ``semicovalent bonding".
However, when $\phi$ decreases, 
the overlap of Mn and O orbitals becomes smaller and Kanamori-Goodenough rules are not sufficient to explain the complex phase diagram of manganites. Instead, the next-nearest-neighbor antiferromagnetic superexchange becomes relatively dominant 
and the E type AFM state becomes stable.\cite{kimuraprb} 
In the following section, we will address quantitatively these general arguments and discuss how magnetic interactions are affected by $\phi$.

\subsection{Structural and magnetic properties}

\subsubsection{\label{sec:struc}Atomic optimization in the AFM-A phase\protect\\}

The atomic structure of \textit{R}MnO$_{3}$ compounds has been investigated by  
neutron powder diffraction \cite{struc_alonso}(for \textit{R}=Pr, Nd, Dy, Tb, Ho, Er, Y), single crystal X-ray diffractometry \cite{struc_mori}(for \textit{R}=Nd, Sm, Eu, Gd) and synchrotron X-ray powder diffraction measurement \cite{struc_tachibana}(for \textit{R}=Ho, Er, Tm, Yb, Lu).

First, in order to discuss the magnetic stability in a tiny range of energy, 
we have used the experimental lattice parameters,
where the volume of the unit cell linearly decreases in the series according to the  Lanthanide contraction upon increasing the atomic number of $R$ atom; 
we have optimized the internal structural parameters imposing AFM-A spin configuration on Mn-$d$ electrons. 
The reason why we used the AFM-A configuration is based on the fact
that most of these compounds (where $R=$ La to Gd) show as ground state the AFM-A spin-configuration.

Second, to discuss ferroelectric properties,
the atomic structure was optimized with AFM-E spin configuration
which, as discussed for HoMnO$_3,$ breaks the inversion symmetry of the system and  leads to a magnetically induced polarization. The results of this second part will be discussed in a later section.

\begin{figure}
\resizebox{78mm}{!}
{
\includegraphics{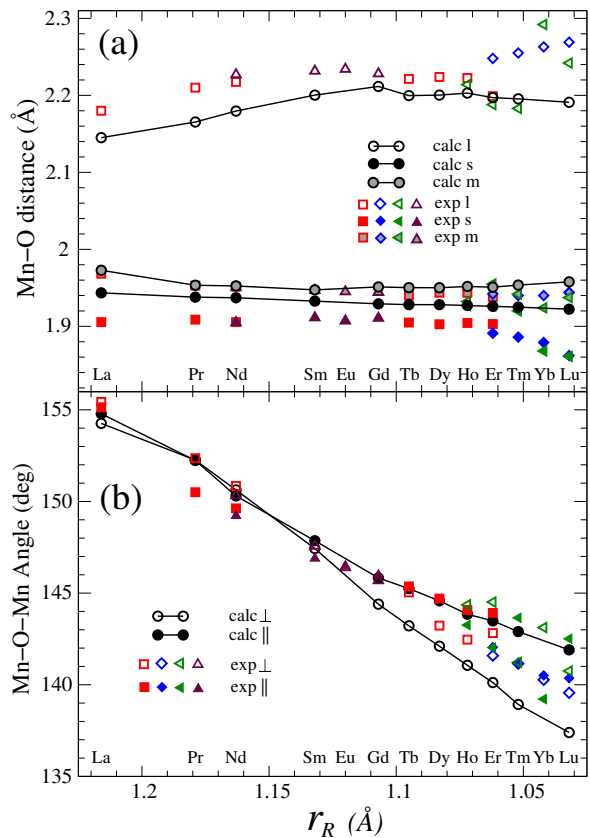}
}
\caption{\label{fig:lms_ang} 
(a) Mn-O bond distance of of \textit{R}MnO$_{3}$ as a function of $r_{R}$ (radius of Rare earth ion $R^{3+}$).
Our results are shown as round symbols: open and filled circles show long and small Mn-O distances in $ac$ plane, respectively, and
the half-filled symbols show a middle Mn-O distance along $b$ axis.
Lines are plotted for an eye guide.
(b) Mn-O-Mn angles of \textit{R}MnO$_{3}$ as a function of the radius of Rare earth atom.
The open symbol shows inter-plane Mn-O-Mn angles (with apical O along $b$ axis) whereas 
the closed symbol shows in the $ac$ plane Mn-O-Mn angles.
In both panels, experimental data are shown  for comparison:
square from \cite{struc_alonso},
diamond from \cite{struc_tachibana},
left triangle from \cite{struc_zhou},
upper triangle from \cite{struc_mori}.
  }
\end{figure}

\begin{table*}
\label{struc.a}
\caption{\label{tab:struc}Structural parameters optimized with AFM-A configuration in $Pnma$ unit cell where the origin is fixed at the  position on Mn atom. 
The lattice parameter used for calculations are obtained 
from Ref.\cite{struc_alonso} for La, Pr, Nd, Tb, Dy and Ho, 
from Ref.\cite{struc_tachibana} for Er, Tm, Yb and Lu, 
from Ref.\cite{struc_mori} for Sm, Eu and Gd.
}
\begin{ruledtabular}
\begin{tabular}{lcccccccccccc}
&La & Pr & Nd & Sm & Gd & Tb & Dy & Ho & Er & Tm & Lu  \\
\hline
a (\AA) & 5.7473 & 5.8129 & 5.8317 & 5.8620 & 5.8660 & 5.8384 & 5.8337 & 5.8354 & 5.8223 & 5.8085 & 5.7868 & \\
b (\AA) & 7.6929 & 7.5856 & 7.5546 & 7.4770 & 7.4310 & 7.4025 & 7.3778 & 7.3606 & 7.3357 & 7.3175 & 7.2959 & \\
c (\AA) & 5.5367 & 5.4491 & 5.4170 & 5.3620 & 5.3180 & 5.2931 & 5.2785 & 5.2572 & 5.2395 & 5.2277 & 5.1972 & \\
\hline
$R$  4$c$ (x $\frac{1}{4}$ z) \\
\ x & 0.0508 & 0.0654 & 0.0714 & 0.0785 & 0.0829 & 0.0836 & 0.0846 & 0.0856 & 0.0864 & 0.0867 & 0.0870 & \\
\ z & 0.4907 & 0.4866 & 0.4849 & 0.4827 & 0.4816 & 0.4808 & 0.4806 & 0.4805 & 0.4798 & 0.4799 & 0.4800 & \\
\hline
Mn 4$a$ (0 0 0) \\
\hline
O1 4$c$ (x $\frac{1}{4}$ z) \\
x & 0.4856 & 0.4820 & 0.4798 & 0.4745 & 0.4683 & 0.4665 & 0.4643 & 0.4617 & 0.4600 & 0.4574 & 0.4536 & \\
z & 0.5779 & 0.5838 & 0.5888 & 0.5979 & 0.6066 & 0.6102 & 0.6133 & 0.6162 & 0.6190 & 0.6223 & 0.6267 & \\
\hline
O2 8$d$ (x y z) \\
x & 0.3014 & 0.3079 & 0.3124 & 0.3181 & 0.3224 & 0.3223 & 0.3235 & 0.3250 & 0.3250 & 0.3258 & 0.3272 & \\
y & 0.0404 & 0.0429 & 0.0452 & 0.0485 & 0.0515 & 0.0530 & 0.0541 & 0.0550 & 0.0561 & 0.0573 & 0.0591 & \\
z & 0.2215 & 0.2155 & 0.2117 & 0.2071 & 0.2033 & 0.2020 & 0.2006 & 0.1988 & 0.1981 & 0.1972 & 0.1952 & \\
\end{tabular}
\end{ruledtabular}
\end{table*}
%
The optimized structural parameters in AFM-A spin configuration, reported in Table.\ref{tab:struc} and Fig.\ref{fig:lms_ang}, give a rather regular trend in the series.
As for the Mn-O bond length, $m$ and $s$ look rather constant over the series, whereas $l$ shows a broad maximum at Gd. Although calculated inter-planar lengths  $m$ show good agreement with corresponding experimental values,
the in-plane lengths  $l$ and $s$ show some deviation ($<$ 10$\%$) with respect to experiments, so that
the JT-distortion is underestimated in our calculations.
The deviation mainly derives from the lack of an exact treatment of the exchange-correlation potential (and/or many-body effects) at the LDA/GGA level: in our simulations,
the rather correlated Mn-$d$ electrons are described as more delocalized than in real manganites, so that the JT-distortion is reduced. \cite{weiku}

As for the Mn-O-Mn bond angle, it decreases almost linearly with the $R$ ionic radius. 
The inter-plane angle becomes larger than in-plane angle for small values of  the radius or $R$ ion. Moreover, the difference between the two Mn-O distances  ($l$ and $s$) becomes rather constant at $R =$ Gd.
This implies that the JT-distortion is rather ``saturated" compared with the GdFeO$_{3}$-like tilting in a Mn-O$_{6}$
octahedron.

\begin{figure}
\resizebox{78mm}{!}
{
\includegraphics{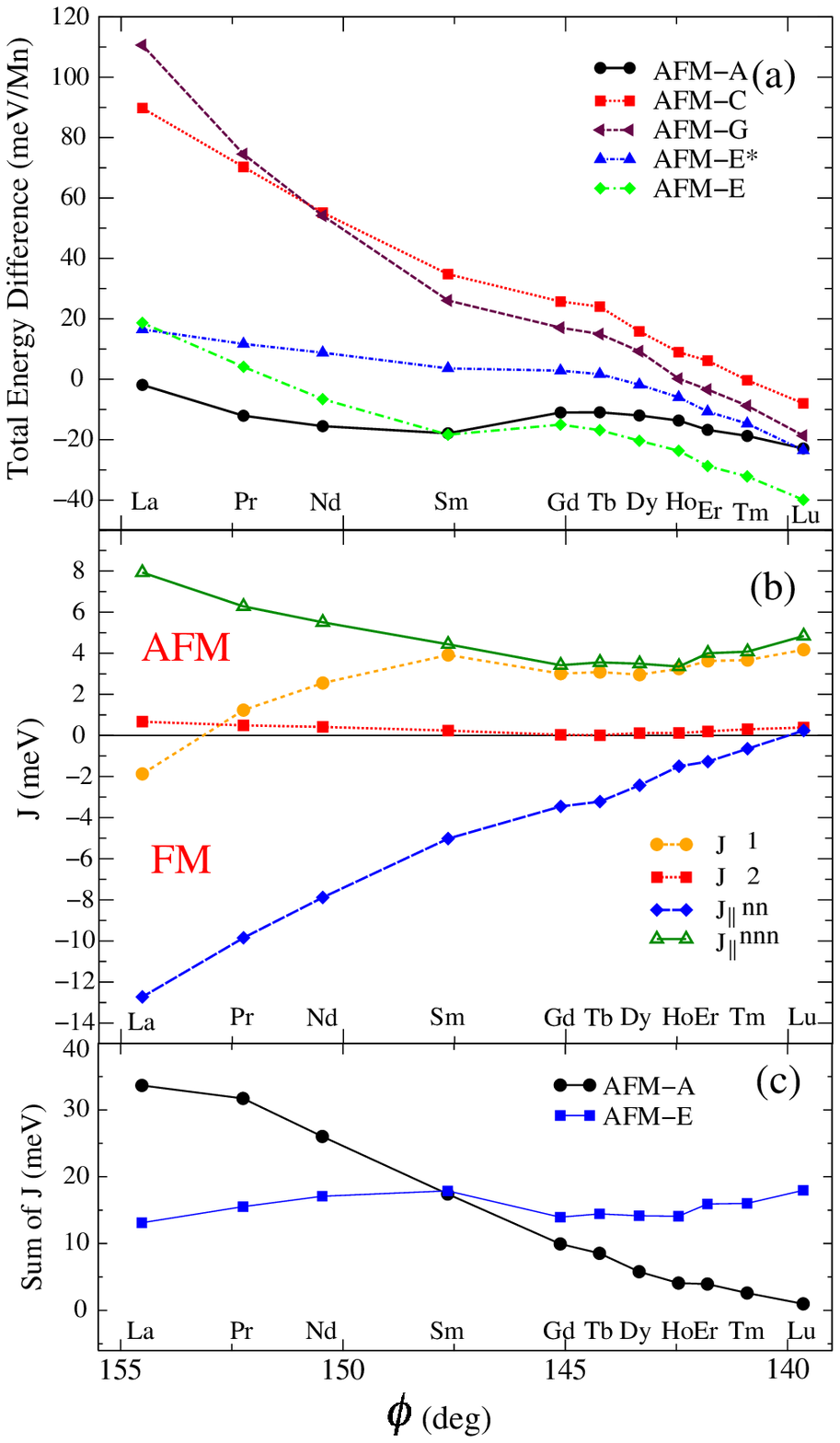}
}
\caption{\label{fig:en_afm} 
(a) Total energy of AFM phases with respect to FM phase.
(b) Super exchange energy ($J_{ij}$):
The first nearest-neighbor in-plane 
$J^\textrm{nn}_{\parallel}$,
next nearest-neighbor in-plane 
$J^\textrm{nnn}_{\parallel}$,
first nearest-neighbor inter plane
 $J^{1}_{\bot}$,
next nearest-neighbor inter plane
 $J^{2}_{\bot}$.
(c) Total energy of AFM-A and AFM-E phase within mean-field approximation ({\em i.e.} sum of $J_{ij}$).
}
\end{figure}
\subsubsection{Magnetic Stability}
Using the optimized atomic structure obtained by imposing the AFM-A spin configuration, 
the total energy of AFM-A, -C, -G, -E and -E* was calculated (Fig.\ref{fig:en_afm}(a)).
This result shows a very clear trend which gives
the transition from AFM-A phase to AFM-E phase at  SmMnO$_{3}$.
Although the optimization imposing AFM-A spin ordering enhances the stability of AFM-A,
the results show that the AFM-E phase is most stable at $R$= Gd,
at variance with  experimental results reporting that the AFM-A phase is the spin ground-state even in GdMnO$_{3}$. The disagreement with experiments may again come from the use of LDA/GGA.
Possible improvements might be obtained upon use of an LDA/GGA+$U$ effective potential\cite{ldau} to enhance the localization of Mn-$d$ state and reproduce the correct ground state in GdMnO$_{3}$.
However, the change of energy is quite sensitive to the value of $U$ (which is, by the way, unknown for most of the compounds):
in our previous calculations \cite{silviamag} ,
small value of $U$ ($\sim 2.5$eV) were found to stabilize AFM-A phase with respect to AFM-E in HoMnO$_{3}$, at variance with experiments which show the AFM-E as ground-state.\cite{munozho,lorenz}
In order to avoid any additional parameters and to discuss the trend of the manganites series without further bias, 
we don't employ the LDA/GGA+U approach and focus on qualitative prediction of chemical trends (as a function of the rare-earth ion), which are expected to be well reproduced within a bare GGA approach.

Using Heisenberg Hamiltonian with normalized spin moment:
\begin{equation}
\mathcal{H}=\sum_{\langle i,j\rangle}{J_{ij}\vec{s_{i}}\cdot\vec{s{_{j}}}/\left|\vec{s_{i}}\right|\left|\vec{s_{j}}\right|}.
\end{equation}
we estimated the super exchange interaction energies $J_{ij}$.
From the following six equations, the difference of total energy between each AFM phase and the reference FM phase is calculated.
Then using a least square mean method, 
we obtained four parameters for $J_{ij}$: the first-nearest-neighbor $J^\textrm{nn}_{\parallel}$ and second-nearest-neighbor coupling along $a$ axis $J^\textrm{nnn}_{\parallel}$ in the $ac$ plane, as well as the first- and second-nearest-neighbor coupling out of plane $J^{1}_{\bot}$ and $J^{2}_{\bot}$. 
With these considered five AFM-configurations, the contribution from second-nearest-neighbor coupling along the $c$ axis is not taken into account and, therefore, cannot be determined.
\begin{eqnarray}
\textrm{FM}&:&       E = 4J^\textrm{nn}_{\parallel}+2J^\textrm{nnn}_{\parallel}+2J^{1}_{\bot}+8J^{2}_{\bot} \\
\textrm{AFM-A}&:& E = 4J^\textrm{nn}_{\parallel}+2J^\textrm{nnn}_{\parallel}-2J^{1}_{\bot}-8J^{2}_{\bot} \label{eq:a}\\
\textrm{AFM-C}&:& E =-4J^\textrm{nn}_{\parallel}+2J^\textrm{nnn}_{\parallel}+2J^{1}_{\bot}-8J^{2}_{\bot}\\
\textrm{AFM-G}&:& E =-4J^\textrm{nn}_{\parallel}+2J^\textrm{nnn}_{\parallel}-2J^{1}_{\bot}+8J^{2}_{\bot}\\
\textrm{AFM-E}&:& E = 					      -2J^\textrm{nnn}_{\parallel}-2J^{1}_{\bot} \label{eq:e}\\
\textrm{AFM-E*}&:& E = 					      -2J^\textrm{nnn}_{\parallel}+2J^{1}_{\bot}
\label{eq:jij}.
\end{eqnarray}
As shown in Fig.\ref{fig:en_afm}(b),
consistently with Kanamori-Goodenough rules,\cite{kanamorigoodenough}
the magnitude of ferromagnetic $J_{||}^\textrm{nn}$ decreases 
with $R$.  $J_{||}^\textrm{nn}$  represents the sum of two competing interactions (FM coupling due to $e_g$ orbitals and AFM coupling due to $t_{2g}$ coupling); the global FM behaviour shows that the
former dominates. However, when $\phi$ decreases, the orbital overlap between Mn-$e_{g}$ orbitals making $\sigma$ bonding with O-$p$ orbital is strongly reduced and this, in turn, reduces the FM character of  
$J_{||}^\textrm{nn}$. 
In-plane anti-ferromagnetic $J_{||}^\textrm{nnn}$ stays rather constant, at variance with what proposed in
the model study by Kimura $et$ $al$  where $J_{||}^\textrm{nnn}$ increases with $\phi$.\cite{kimuraprb}
The interplane coupling is expected to be AFM, due to its mainly $t_{2g}$-driven character. The fact that it is estimated to be weakly FM ($<$ 2 meV) in LaMnO$_3$ casts some doubts about the validity of a bare GGA treatment for LaMnO$_3$ (where the AFM-A spin-state is found basically degenerate with the FM spin configuration, obviously at variance with experiments). On the other hand, the correct spin ground-state is reproduced for most of the $R$MnO$_3$
(with exceptions of {\em i}) the above mentioned Gd and{\em ii}) Tb and Dy where we did not attempt to simulate the non-collinear spiral arrangements, due to further complexity in the simulations).\cite{kimuraprb}

Figure \ref{fig:en_afm}(c) shows the total energy using Eqn.(\ref{eq:a}) and (\ref{eq:e}): within the mean field approximation,
this  energy is supposed to be proportional to the ordering Neel temperature $T_\textrm{N}$. Indeed, the trend 
is in good agreeement with experimental results, 
showing a steeply decreasing $T_\textrm{N}$ with AFM-A in the first half of the series and  a rather constant $T_\textrm{N}$ with AFM-E in the second half. 

As a summary of this section,  an
increase of $\phi$ strongly reduces $e_{g}$-derived $J_{||}^\textrm{nn}$ but 
doesn't change drastically the  $J_{||}^\textrm{nnn}$ exchange constant,
so that the AFM-A phase shows a spin transition to AFM-E phase. 

\subsection{The multiferroic AFM-E spin configuration}

\subsubsection{Structural properties}
\begin{figure}
\resizebox{78mm}{!}
{
\includegraphics{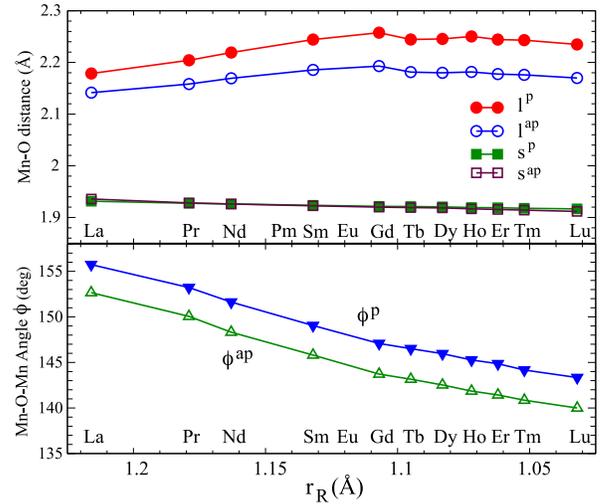}
}
\caption{\label{fig:opt_e.ls_ang} The Mn-O bond length (upper panel) and Mn-O-Mn bond angle (lower panel) as a function of $r_{R}$ in $R$MnO$_{3}$ structure optimized in AFM-E phase.
Here, the superscript $p$ ($ap$) denotes parallel (antiparallel) spin of Mn ions.
}
\end{figure}
In order to discuss the chemical trends of ferroelectric properties,
we optimized the internal atomic positions in the $R$MnO$_{3}$ systems by artificially imposing the AFM-E spin configuration for all $R$MnO$_{3}$ (irrespective of whether or not the AFM-E is the spin ground-state).
The resulting symmetry (lowered by the spin configuration with respect to the AFM-A spin arrangements), allows two
significantly different values of $l$ and $\phi$ between parallel and anti-parallel Mn-spins  whereas the optimized $s$ doesn't change (cfr Fig.\ref{fig:opt_e.ls_ang}).
This is consistent with our mechanism proposed for HoMnO$_3$ in the previous section:
since the long Mn-O bond length is mostly affected by the ferroelectric AFM-E spin configuration (cfr Fig.\ref{fig:opt_e.ls_ang}) and it is the one along which the Mn-$d_{z^{2}}$ orbital is pointing, it is likely that this latter orbital plays a key role in the final ferroelectricity.
We also note that, along the series, the difference between $\phi^p$ and $\phi^{ap}$ stays rather constant; since, according to Ref. \cite{ivan}, this difference is at the basis of the polar atomic displacements, this constant behaviour will be relevant in the discussion of polarization trends vs $R$ (see below).%

\subsubsection{Electronic States and Orbital Ordering}
\begin{figure}
\resizebox{78mm}{!}
{
\includegraphics{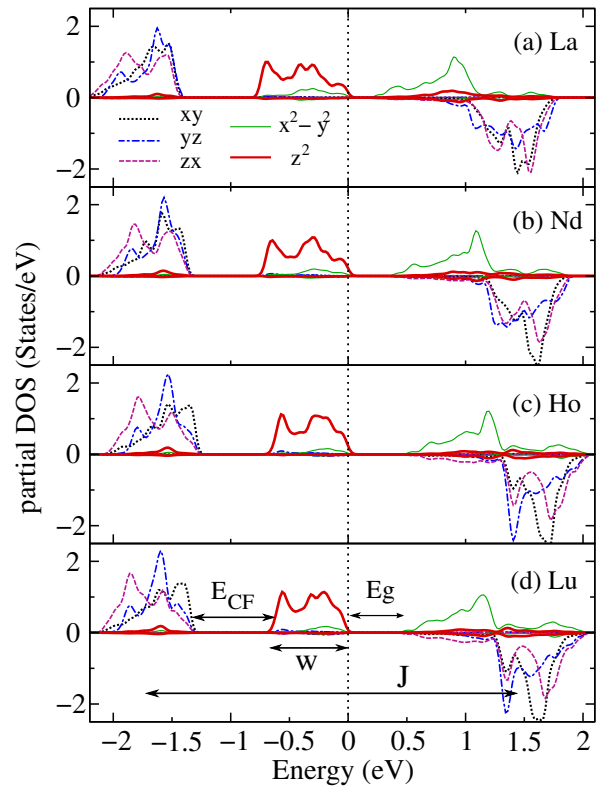}
}
\caption{\label{fig:pdos} The orbital-decomposed partial DOS of Mn-3$d$ components for $R$MnO$_{3}$ : $R$ = a) La, b) Nd, c) Ho, d) Lu. 
Arrows indicate: Energy gap ($E_{g}$), crystal field splitting ($E_{\textrm{CF}}$), $e_{g}$ bandwidth ($w$) and on-site exchange interaction energy ($J$).
The crystal structure was optimized imposing the AFM-E spin configuration.
The local frame in the MnO$_{6}$ octahedron is chosen as x:middle(inter), y:short, z:long axis.}
\end{figure}
In order to better understand the trends of the ferroelectric properties as a function of $R$, reported in the following section, let us briefly discuss the electronic structure of the compounds in the AFM-E spin configuration. 

As far as the magnetic moment in the Mn muffin-tin sphere is concerned, it is basically constant and equal to $\sim$ 3.31 $\mu_B$ for all $R$. However, some differences arise when looking at the partial density of states (pDOS) for Mn-$3d$ state along the series (projected in the ``local" octahedron frame to highlight orbital-ordering), as reported in 
Figure \ref{fig:pdos}.
It is clearly evident that the d$_{z^{2}}$ and d$_{x^{2}-y^{2}}$ states are fully orbitally-polarized.
More quantitatively, the coefficient of the d$_{z^{2}}$ orbital, obtained by diagonalization of density matrix for five $d$ orbitals, is 0.98 at $R$=La and 0.96 at $R$=Lu. This
implies  that the orbital ordering is already ``saturated" at LaMnO$_{3}$.

A progressive distortion of the structure is shown to increase the
energy gap $E_{g}$ and to decrease the width of the $e_{g}$-band ($w$). 
Within a  tight-binding framework, $w$ is proportional to the hopping integral $t$. Indeed, we estimated the hopping integral for selected compounds along the series and found that $t$ increases with the ionic radius of the $R$ ion (see Table \ref{tab:hop}).

\begin{table}
\caption{\label{tab:hop}
Hopping integral  between Mn-$e_{g}$ WFs with same spin state in optimized AFM-E $R$MnO$_{3}$ (meV).}
\begin{ruledtabular}
\begin{tabular}{lcccccccccccc}
    &$R=$ La     &Nd    &Sm    &Ho    &Lu\\
\hline
$t$  &128   &128  &124  &117  &112\\
\end{tabular}
\end{ruledtabular}
\end{table}

Here, we would like to point out that the DOS doesn't change significantly before and after FE atomic displacements, at variance with standard FE (such as BaTiO$_3$), where FE atomic displacements are accompanied by a rehybridization of filled O-$p$ and empty cation $d$ states.\cite{nicolabto}
Here,  we recall that the Mn-$d$ state is well localized and the character of the Mn-O bond is rather ionic in nature so it is not expected to undergo drastic changes upon development of polarization as far as the bonding properties are concerned.

\subsubsection{Ferroelectricity and link to lattice, spin and orbital degrees of freedom}

\begin{figure}
\resizebox{78mm}{!}
{
\includegraphics{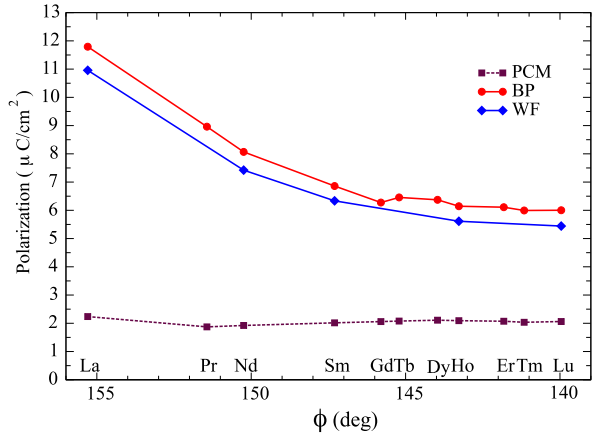}
}
\caption{\label{fig:berry} Ferroelectric polarization calculated by Berry phase method (BP), Wannier function method (WF) and point charge model (\textrm{PCM}) as a function of $\phi$ in AFM-E $R$MnO$_{3}$ (see text for details). The in-plane Mn-O-Mn bond angle $\phi$ is averaged between $\phi_{p}$ and $\phi_{ap}$.
}
\end{figure}

\begin{table}
\caption{\label{tab:wfc}
Polarization ($\mu$C/cm$^{2}$) induced by the displacement of Wannier function center (WFC) from the ionic position in AFM-E $R$MnO$_{3}$. Only up-spin contribution is reported for each WF: Mn-$e_g$, Mn-$t_{2g}$ and O-$p$. In the two final lines, we show the sum of the contributions (for spin-up only) and the total (for spin-up and spin-down).}
\begin{ruledtabular}
\begin{tabular}{lcccccccccccc}
 $P$        &$R=$ La     &Nd    &Sm    &Ho    &Lu\\
\hline
Mn-$e_{g}$ &-4.76&-3.48&-3.01&-2.67&-2.61\\
Mn-$t_{2g}$& 1.36& 1.73& 2.02& 2.42&2.47\\
O-$p$	     &-0.96&-1.00&-1.17&-1.51&-1.55\\
\hline
total (up) &-4.36&-2.75&-2.16&-1.76&-1.69\\ \hline
total (up+down) &-8.72&-5.50&-4.32&-3.52&-3.38\\
\end{tabular}
\end{ruledtabular}
\end{table}

Similarly to the case of HoMnO$_3$, we have calculated the polarization according to the PCM, BP and WF approaches.
As shown in Fig.\ref{fig:berry}, the polarization calculated by BP and WF are consistent within less than 10 \%: this difference (in addition to possible numerical uncertainties deriving from the use of a different 
basis set and potential treatments in VASP and FLEUR codes for BP and WF approaches, respectively)
comes from O-$2s$,  $R$-$5s$, $R$-$5p$ contributions and shows that these latter contributions, as expected, are small compared to Mn-$d$ and O-$p$ contributions. 

What is truly remarkable is that the polarization (both from BP and WF) shows a rapid increase upon decreasing of $\phi$ ($P_{\textrm{BP}}$  approaches 12 $\mu C/cm^2$ in an hypothetical AFM-E LaMnO$_3$), whereas $P_{\textrm{PCM}}$ is to a large extent constant. This latter trend is consistent with the constant behavior of $\phi^p-\phi^{ap}$ shown in Fig.\ref{fig:opt_e.ls_ang}  and with what previously discussed in Refs. \cite{ivan,prlslv}.
Moreover, the difference between P$_\textrm{BP}$/P$_\textrm{WF}$ and P$_{\textrm{PCM}}$ (which represents a purely electronic contribution)
is much bigger than the contribution coming from atomic displacementsin the entire series. 
This confirms the different nature of the magnetic origin of ferroelectricity in manganites with respect to standard ``proper" ferroelectrics. 

Our results further suggest that  orbital ordering is necessary for the rising of the polarization; however, being constant along the series, 
is not responsible for the {\em trend} of FE polarization, which, on the other hand, shows dramatic changes as a function of $R$.
Consistently with what previously discussed for HoMnO$_3$ where $t$  
is suggested to play a important role, the dramatic increase in the {\em asymmetric} $e_g$ hopping is responsible for enhancing 
$P$ in less distorted manganites (cfr Tab. \ref{tab:wfc} where we report the different contribution of Mn-$e_g$, Mn-$t_{2g}$ and O-$p$ to the total $P_\textrm{WF}$). We show that, for every $R$, the Mn-$e_g$ and O-$p$ contributions show an opposite sign with respect to the Mn-$t_{2g}$ term with regular trends along the series.
 Indeed, remembering that the hopping integral $t$ depends on the bond length and bond angle, 
one expects smaller $l$ and larger $\phi$ to enhance $t$ between $e_{g}$ orbitals, 
whereas smaller $s$ to enhance $t$ between $t_{2g}$ orbitals; this is confirmed  by looking carefully at table.\ref{tab:wfc} and Fig.\ref{fig:opt_e.ls_ang} and their implications for the values of P (cfr. Fig. \ref{fig:berry}).

\section{conclusions}

Ferroelectricity, recently proposed for the collinear AFM-E magnetic phase in orthorhombic
HoMnO$_3$, is here explained microscopically from first-principles via a careful Wannier function analysis. We show that the {\em asymmetric} electron hopping of orbitally-polarized Mn-$e_g$ states is the key ingredient for the rising of polarization. At variance with proper ferroelectrics,  in HoMnO$_3$ the purely electronic contribution (due to Wannier function centers which are largely displaced with respect to ions) dominates the polarization,
with respect to the contribution to $P$ coming from ionic movements. However, the net polarization along the $c$ axis is the result of  a delicate balance of different contributions, such as, for example, the opposite signs of the  Mn $e_g$ and $t_{2g}$ contributions to $P$.

In addition, extensive {\em ab--initio} calculations have been performed for the $R$MnO$_3$ systems, focusing on the link between ferroelectricity and the spin, orbital and lattice degrees of freedom in the aim of identifying chemical trends along the series.
In summary, the results are summarized as:
{\em i}) ferromagnetic $J^\textrm{nn}_{\parallel}$ increases with the Mn-O-Mn angle $\phi$, whereas antiferromagnetic $J^\textrm{nnn}_{\parallel}$ is rather constant with $\phi$;
{\em ii}) $P_{e_{g}}$ increases with $\phi$ and decreases with $l$, whereas
 $P_{t_{2g}}$ decreases with $s$ and partially cancels $P_{e_{g}}$. 
{\em iii}) orbital ordering, a needed ingredient in the rising of $P$, is however saturated along the series and does not influence the trend of polarization as a function of $R$.

Therefore, the main message is that one can expect high ferroelectric polarization with large $\phi$ and small $l$ in the AFM-E phase;
however, such large $\phi$ would simultaneously increase the ferromagnetic 
$J^\textrm{nn}_{\parallel}$, therefore stabilizing the centrosymmetric AFM-A spin configuration instead of the polar AFM-E phase.
The problem is then
{\em how}  to increase $P$, still keeping the AFM-E phase as stable spin-state ({\em i.e.} via strain, pressure, alloying, etc):
we hope that our findings will be helpful to answer this question. 
\acknowledgments
We thank Elbio Dagotto and Claude Ederer for  their careful reading of the manuscript.
K.Y. acknowledges support from bilateral agreement between Consiglio Nazionale delle Ricerche and Deutsche Forschungsgemeinschaft (DFG). Computational support from Barcelona Supercomputing Center is gratefully acknowledged. 
In this paper, the crystal structure and WFs are plotted using the program XCrySDen\cite{xcrysden}.


\begin{thebibliography}{100}
\bibitem{notaberry} We recall that the polarization is obtained as a vectorial sum of the displacement of the Wannier function centers with respect to the ionic positions + the contributions due to the point-charge-model
\bibitem{notaperp} In our previous calculations \protect\cite{prlslv}, our reference paraelectric structure was different from AFM-A chosen here and showed some non-collinear spin-arrangements. However, the different path does not affect the final results as far as the polarization is concerned.
\end{thebibliography}

\end{document}